\documentstyle[12pt]{article}
\begin{document}
\vskip .7cm
\begin{center}
{\large{\bf { Hodge decomposition theorem \\for Abelian
 two-form gauge theory}}}
\vskip 2cm

{\bf E. Harikumar$^{*}$, R. P. Malik$^{{**},\dagger}$ and M. Sivakumar
$^{{*}, \ddagger}$}
\vskip 0.5cm
$^{*}$ {\it School of Physics, University of Hyderabad,}\\
{\it Hyderabad-500 046, India}\\
$^{**}$ {\it S. N. Bose National Centre for Basic Sciences,} \\
{\it Block-JD, Sector-III, Salt Lake, Calcutta-700 098, India}\\

\vskip 2.5cm

\end{center}

\noindent
{\bf Abstract}.
We show that the BRST/anti-BRST  invariant  $3+1$ dimensional 2-form  
gauge theory has further nilpotent symmetries (dual BRST /anti-dual BRST) that 
leave the gauge fixing term invariant. The generator 
for the dual BRST symmetry is analogous to the co-exterior derivative
of differential geometry. There exists a bosonic symmetry which keeps 
the ghost terms invariant and it turns out to be the analogue
of the Laplacian operator. The Hodge duality operation is shown to correspond
to a discrete symmetry in the theory.  The generators of all these continuous 
symmetries are shown to obey the algebra of the de Rham cohomology operators of 
differential geometry. We derive the extended BRST algebra  constituted by six 
conserved charges and discuss the Hodge
decomposition theorem in the quantum Hilbert space of states.

\baselineskip=16pt

\footnotetext{$^{\dagger}$ E-mail: malik@boson.bose.res.in}
\footnotetext{$^{\ddagger}$ E-mail: mssp@uohyd.ernet.in}

\newpage

\noindent
{\bf 1 Introduction}\\

\noindent
For the covariant canonical quantization of gauge theories, one of the most
elegant methods is the Becchi-Rouet-Stora-Tyutin (BRST) formalism [1, 2] where
(quantum) gauge invariance and unitarity are respected {\it together} at any
arbitrary order of perturbation theory. The first-class constraints of the
original gauge theories are found to be encoded in the subsidiary condition
($ Q_{B} |phys> = 0$) when one requires that the physical subspace (of the
total Hilbert space of states) contains only those states that are annihilated 
by the nilpotent ($Q_{B}^2 = 0$) and conserved ($\dot Q_{B} = 0$)
BRST charge $Q_{B}$. 
In fact, the condition $Q_{B} |phys> = 0$ implies that the operator 
form of the first-class constraints
annihilate the physical states. This requirement is essential for the 
consistent quantization
of any theory endowed with the first-class constraints (Dirac's prescription) 
[3, 4].
The nilpotency of the BRST charge ($Q_{B}^2 = 0$) and physicality criteria
($Q_{B} |phys> = 0$) are the two key requirements for the discussion of
cohomological aspects of BRST formalism [5-8] and   its connection with
the de Rham cohomology operator $d$ (exterior derivative; $d^2 = 0$)
of differential geometry defined
on a compact manifold. For instance, two physical
states are said to belong to the same cohomology class w.r.t. $Q_{B}$
if they differ by a BRST exact state
as two closed forms belong to the same cohomology class w.r.t. operator $d$
if they differ by an exact form. There are two other de Rham cohomology
operators that are essential for the definition of the Hodge decomposition
theorem which states that, on a compact manifold, any arbitrary $n$-form
$f_{n} (n = 0, 1, 2, 3.....)$ can be written as a unique sum of a harmonic 
form $ h_{n}
(\Delta h_{n} = 0, d h_{n} = 0, \delta h_{n} = 0) $, an exact 
form ($ d e_{n-1}$) and  a co-exact form ($\delta c_{n+1}$):
$$
\begin{array}{lcl}
f_{n} = h_{n} + d\; e_{n-1} + \delta\; c_{n+1}
\end{array} \eqno(1.1)
$$
where $ \delta = \pm * d *, (\delta^2 = 0) $ is the dual exterior derivative,
 $ \Delta = (d + \delta)^2 = d \delta + \delta d$ is the Laplacian 
operator and $*$ is the so-called Hodge duality operation [9-12].

It is a well known fact that the  cohomological operator $d$ of 
differential geometry finds its analogue in the local, conserved 
and nilpotent BRST charge $Q_{B}$ [7, 8]. 
It is, therefore, interesting to enquire if analogous local conserved 
charges (and corresponding local symmetry transformations for a given
Lagrangian density) exist for the
analogues of the other cohomological operators, viz; $\delta$ and $\Delta$.
Some interesting attempts [13-16] have been made to express $\delta$ and
$\Delta$ in the language of symmetry properties of a given Lagrangian
density for the 1-form interacting gauge theory in any arbitrary 
spacetime dimension.
The symmetry transformations,
however, turn out to be non-local and non-covariant. In the covariant 
formulation [17], the nilpotency of transformations are dependent on the 
specific choice of  parameters of the theory. Recently,  it has been 
shown that the two-dimensional (2D)  free Abelian as well as non-Abelian
gauge theories (without any interaction with matter fields) provide a
{\it topological}
\footnote{ A theory with a flat spacetime metric and
without any propagating degrees of freedom.}
field theoretical model for the Hodge theory where
symmetry transformations corresponding to the de Rham cohomology operators
($ d, \delta, \Delta$) are nilpotent (for $d$ and $\delta$), local,
covariant and continuous [18-20]. The analogue of these local symmetries
have also been shown to exist for the 2D topological fields  
(i.e., 2D Abelian gauge fields) coupled to matter (Dirac) fields
in two-dimensions of spacetime [21].

In our present paper, we show the existence of symmetries 
corresponding to the de Rham cohomology operators for a field
theoretical model in the physical four ($ 3 + 1)$ dimensional spacetime
\footnote{ We follow the notations in which the flat Minkowski metric is
$\eta_{\mu\nu} =$ diag $ (+, -, -, -)$ and Levi-Civita totally
antisymmetric tensor $\varepsilon_{0123} = + 1 = - \varepsilon^{0123},\;
\varepsilon_{0ijk} = \varepsilon_{ijk} = - \varepsilon^{ijk},\;
\varepsilon^{\mu\nu\lambda\xi} \varepsilon_{\mu\nu\lambda\xi} = - 4!,\;
\varepsilon^{\mu\nu\lambda\xi} \varepsilon_{\mu\nu\lambda\rho}
= - 3! \;\delta^{\xi}_{\rho}, $ etc.
Here Greek indices: $\mu, \nu, \lambda.......$ =
0, 1, 2, 3 and Latin indices: i , j, k .......= 1, 2, 3.}. 
The search for such symmetries in the Abelian and non-Abelian 1-form gauge 
theories, even though quite illuminating,
has not been fully successful and satisfactory, as stated earlier.
Thus, the central theme of our present work
is to show that the free Abelian antisymmetric (2-form) gauge theory in 4D
provides a prototype example for Hodge theory where the de Rham cohomology
operators correspond to the local and conserved charges. These charges 
turn out to be the generators of specific {\it local}, {\it covariant}
and {\it continuous} symmtery transformations 
for the BRST invariant Lagrangian density of this theory.

The 2-form massless gauge theory is interesting by itself as it is a 
dual description for the massless scalar theory.  
It also has interesting constraint structure: stage-one reducibility and
corresponding ghost for the ghost feature. In addition, the 2-form potential
also appears naturally in supergravity and superstring theories including the
recent developments in non-commutative geometry [22].
Its different forms have appeared in other 
contexts of theoretical physics, e.g., QCD, cosmic strings and vortices,  
black holes, etc. [23-26]. In fact, this theory,
coupled to a 1-form Abelian gauge field via a `topological' 
$ B \wedge F$  term, has rich mathematical structure and has been studied 
from various points view, viz., duality consideration [27,28], Dirac
bracket analyses [29,30,31], BFT Hamiltonian formulation [32], BRST 
quantization [33], etc.

We shall consider the BRST invariant version (see, e.g., Section 2 (below))
of the free 4D Kalb-Ramond Lagrangian density [23,34,35] 
$$
\begin{array}{lcl}
{\cal L} = \frac{1}{12}\; H^{\mu\nu\lambda} H_{\mu\nu\lambda}
\end{array} \eqno(1.2)
$$
where $H_{\mu\nu\lambda} = \partial_{\mu} B_{\nu\lambda} +
\partial_{\nu} B_{\lambda \mu} + \partial_{\lambda} B_{\mu\nu} $ is the
totally antisymmetric curvature tensor constructed from the antisymmetric
gauge field $B_{\mu\nu}$
\footnote{ The  gauge field $B_{\mu\nu}$ is defined through
2-form: $ B = \frac{1}{2} B_{\mu\nu} dx^\mu \wedge dx^\nu$ and
the curvature tensor $H_{\mu\nu\lambda}$ is defined through 3-form
as: $H = d B$. It can be readily seen that the gauge-fixing term
$\partial_{\mu} B^{\mu\nu}$ can be defined through one-form by 
the application of $\delta$ as:
$\partial_{\mu} B^{\mu\nu} dx_{\nu} = \delta B$ where $\delta = - * d *$
is the dual exterior derivative of $d$. It
is clear that the gauge-fixing term is the `Hodge dual'
of curvature term.} 
and show that: 
(i) in addition to the usual BRST charge ($Q_{B}$), there exists a local, 
conserved and nilpotent dual(co)-BRST charge ($Q_{D}$) under
which the gauge-fixing term of this theory remains invariant. This fact should
be contrasted with the usual BRST transformations, under which, 
it is the kinetic energy term (more precisely curvature tensor 
$H_{\mu\nu\lambda}$ itself) that remains invariant. 
(ii)  The anticommutator of BRST- and dual BRST transformations leads to a 
symmetry transformation that is generated by a local and conserved bosonic 
charge ($W$). This is analogous to the Laplacian operator in differential 
geometry where it is given by the anticommutator of $d$ and $\delta$.
(iii) The conserved charges (e.g., $Q_{B}, Q_{D}, W$)
can be exploited together for the discussion
of the Hodge decomposition theorem in the quantum Hilbert space of states and
for the anlaysis of the constraint structure on the 
physical (harmonic) states of the theory. 
(iv) A discrete transformation symmetry of the Lagrangian density
 relates $Q_{B}$ and $Q_D$ like a dual symmetry:
 $Q_{B} \rightarrow \; Q_{D},~ Q_{D} \rightarrow \; - Q_{B}$ 
and $ W \rightarrow\; -W$. This relationship maintains the anticommutator
between $Q_B$ and $Q_D $ and the
underlying discrete symmetry turns out to be a
realization of the Hodge $*$ operation of  differential geometry for this 
gauge theory. To the best of our knowledge, this is the first example of a
field theoretical model for the Hodge theory in 
four ($3+1$) dimensional spacetime
where the conserved charges corresponding to the de Rham cohomology operators 
generate the local, continuous  and covariant transformations for the fields.
The existence of new symmetries (corresponding to $\delta$ and $\Delta$)
and their generalizations might turn out to be useful in the proof of 
renormalizability of an {\it interacting} theory where 
the gauge fields are coupled to matter fields. Thus, our
present work is the first step towards our main goal of  having
a complete understanding of the interacting theory.

The outline of our paper is as follows. In section 2, we recapitulate the 
essentials of the BRST formalism for the 2-form gauge theory and set
up the notations for our further discussion. This is followed by the
discussion and derivation of the dual BRST symmetry in section 3. We
derive the symmetry generated by the Casimir operator in section 4 and
obtain the corresponding conserved charge. Section 5 is devoted to the 
derivation of the extended BRST algebra and 
a brief discussion is provided for its possible connection to 
the de Rham cohomology operators of the differential geometry.  In section
6, we discuss the Hodge decomposition theorem in the quantum Hilbert space 
of states and analyze the structure of constraints on the physical 
states of the theory. Finally, in section 7, we make some concluding 
remarks and point out some directions that can be pursued in the future.\\

\noindent
{\bf 2 Preliminary: BRST symmetry }\\

\noindent
We begin with the BRST invariant Lagrangian density [7]
$$
\begin{array}{lcl}
{\cal L}_{B} &=& \frac{1}{12}\; 
H^{\mu\nu\lambda}\; H_{\mu\nu\lambda} 
- \frac{1}{2} B^\mu B_{\mu} 
+ B_\nu \bigl ( \partial_\mu B^{\mu\nu} - \partial^\nu \phi_{1} \bigr )
- \partial_\mu \bar \beta \;\partial^\mu  \beta \nonumber\\
&+& \bigl ( \partial_\mu \bar C_\nu - \partial_\nu \bar C_\mu \bigr ) \;
\partial^\mu C^\nu + \rho \;\bigl ( \partial_\mu C^\mu + \lambda \bigr )
+ \bigl ( \partial_{\mu} \bar C^{\mu} + \rho \bigr )\; \lambda
\end{array}\eqno(2.1)
$$
where $B_\mu,~\phi_{1},~\lambda$ and $\rho$  are the auxiliary fields 
\footnote{ By integrating out the auxiliary fields, we will obtain 
the Lagrangian density which respects the on-shell nilpotent BRST symmetry.}
introduced to have the off-shell nilpotent BRST invariance. 
The following BRST transformations
$$
\begin{array}{lcl}
$$
&& \delta_{B} B_{\mu\nu} = \eta \bigl ( \partial_\mu C_\nu -
\partial_\nu C_\mu \bigr ) \quad
\delta_{B} C_\mu = \eta \partial_\mu \beta \quad
\delta_{B} \bar C_{\mu} = \eta B_{\mu} \nonumber\\
&&\delta_{B} \phi_{1} = - \eta \;\lambda \qquad\;\;\;
 \delta_{B} \bar \beta = \eta \;\rho \qquad\;\;\;
\delta_{B} \bigl ( B_{\mu}, \rho , \lambda, \beta 
\bigr ) = 0
\end{array} \eqno(2.2)
$$
leave the Lagrangian density invariant up to a total derivative term.

The continuous symmetry transformations (2.2) lead to the following nilpotent 
($Q_{B}^2 = 0$) and 
conserved ($ \dot Q_{B} = 0$) BRST charge due to Noether theorem
$$
\begin{array}{lcl}
Q_{B} = {\displaystyle \int} 
d^3 x \;\bigl [\; H^{0ij} \partial_{i} C_{j} + B_{0} \lambda
- \rho \dot \beta + (\partial^{0} C^{i} - \partial^{i} C^{0}) B_{i}
- (\partial^{0} \bar C^{i} - \partial^{i} \bar C^{0}) 
\partial_{i} \beta \;\bigr ]. 
\end{array} \eqno(2.3)
$$
This charge turns out to be the generator for 
the transformations (2.2) if we exploit
the following general relationship
$$
\begin{array}{lcl}
\delta_{B} \Phi = - i \eta \;\bigl [\; \Phi,  Q_{B} \;\bigr ]_{\pm}
\end{array} \eqno(2.4)
$$
where $ [\;,\;]_{\pm}$ stands for the (anti)commutator for the generic field
$\Phi$ being (fermionic)bosonic in nature. For the verification of (2.4),
one has to use the canonical (anti)commutators for the Lagrangian density
(2.1) as given below (with $\hbar = c = 1$)
$$
\begin{array}{lcl}
&& [ B_{0i} (x,t), B^j (y,t) ] = i \delta_{i}^{j} \delta (x - y) \;\qquad\;
[ \beta (x,t), \dot {\bar \beta} (y,t) ] = - i \delta (x -y) \nonumber\\
&& [\; B_{ij} (x,t), \;H^{0kl} (y,t) ]
= i \;(\;\delta_{i}^{k} \;\delta_{j}^{l} - \delta_{i}^l\; \delta_{j}^k\;) 
\;\delta (x -y) \nonumber\\
&& [ \phi_{1} (x,t), B_{0} (y,t) ] = - i \delta (x -y) \qquad\;\;\;\;
[ \bar \beta (x,t), \dot  \beta (y,t) ] = - i \delta (x -y) \nonumber\\
&&\{ C_{0} (x,t), \rho (y,t) \} = - i \delta (x -y) \qquad\;\;\;\;
\{ \bar C_{0} (x,t), \lambda (y,t) \} = i \delta (x -y) \nonumber\\
&& \{ C_{i} (x,t), \Pi^{j}_{C} (y,t) \} = i \delta_{i}^{j} \;\delta (x -y)
\qquad
\{ \bar C_{i} (x,t), \Pi^{j}_{\bar C} (y,t) \} = i \delta_{i}^{j}
\;\delta (x - y)
\end{array}\eqno(2.5)
$$
where $\delta (x -y) $ is the Dirac- delta function in 3D of space 
(i.e., $\delta^{(3)} ({\bf x} - {\bf y})$) and
the expression for the canonical momenta are:
$$
\begin{array}{lcl}
\Pi^{(i)}_{(C)} = - (\partial^0 \bar C^i - \partial^i \bar C^0) \quad
\Pi^{(i)}_{(\bar C)} = (\partial^0  C^i - \partial^i  C^0). 
\end{array}\eqno(2.6)
$$
All the rest of the (anti)commutators are zero.

It can be readily seen that the ghost part of the Lagrangian density
has the following discrete symmetry invariance
$$
\begin{array}{lcl}
&& \beta \rightarrow \mp i \bar \beta \qquad
C_{\mu} \rightarrow \pm i \bar C_{\mu} \qquad
\rho \rightarrow \pm i \lambda \nonumber\\
&& \bar \beta \rightarrow \pm i \beta \qquad
\bar C_{\mu} \rightarrow \pm i C_{\mu} \qquad
\lambda \rightarrow \pm i \rho.
\end{array} \eqno(2.7)
$$
As a result of this
symmetry, one can define an anti-BRST charge $Q_{AB}$ from
(2.3) and one can obtain anti-BRST symmetry from (2.2)
by exploiting (2.7). 
Furthermore, the total Lagrangian density (2.1) remains invariant
under the following transformations:
$$
\begin{array}{lcl}
&& B_{\mu\nu} \rightarrow B_{\mu\nu} \qquad \;\;\phi_{1} \rightarrow \phi_{1}
\qquad \;\; \;\;\;\;B_{\mu} \rightarrow B_{\mu} \nonumber\\
&& \beta \rightarrow e^{2\;\Sigma} \;\beta \qquad\;\;
C_{\mu} \rightarrow  e^{\Sigma} \;C_{\mu} \qquad\;
\lambda \rightarrow  e^{\Sigma}\; \lambda \nonumber\\
&& \bar \beta \rightarrow e^{-2\;\Sigma} \;\bar \beta \qquad
\bar C_{\mu} \rightarrow  e^{-\Sigma} \;\bar C_{\mu} \qquad
\rho \rightarrow  e^{-\Sigma}\; \rho 
\end{array} \eqno(2.8)
$$
where $\Sigma$ is a global (spacetime independent) 
scale transformation parameter. This continuous
symmetry leads to the derivation of a conserved ghost charge ($Q_{g}$) as
$$
\begin{array}{lcl}
Q_{g} = {\displaystyle \int }
d^3 x \;\bigl [ \;C_{i} \Pi^{(i)}_{(C)} + C_{0} \Pi^{(0)}_{(C)}
+ 2 \beta \Pi_{\beta} - 2 \bar \beta \Pi_{\bar \beta} - \bar C^{0} 
\Pi^{(0)}_{(\bar C)} - \bar C_{i} \Pi^{(i)}_{(\bar C)} \;\bigr ]
\end{array} \eqno(2.9)
$$
where $\Pi$s are the canonical momenta w.r.t. ghost fields
\footnote{ Besides (2.6), the other canonical momenta
are $\Pi_{\beta} = - \dot {\bar \beta}, 
\Pi_{\bar \beta} = - \dot \beta, \Pi_{(C)}^{(0)} = - \rho,
\Pi_{(\bar C)}^{(0)} = \lambda. $}. It can be 
readily seen, by exploiting the canonical (anti)commutators of (2.5), that
$$
\begin{array}{lcl}
&& Q_{B}^2 = \frac{1}{2} \{ Q_{B}, Q_{B} \} = 0 \qquad \;\;\;
Q_{AB}^2 = \frac{1}{2} \{ Q_{AB}, Q_{AB} \} = 0 \nonumber\\
&& \{ Q_{B}, Q_{AB} \} = 0 \qquad i [ Q_{g}, Q_{B} ] = + Q_{B} \qquad
i [ Q_{g}, Q_{AB} ] = - Q_{AB}.
\end{array} \eqno(2.10)
$$
Thus, we note that $Q_{B}$ and $Q_{AB}$ are the nilpotent operators
of order 2 (i.e., $Q_{B}^2 = Q_{AB}^2 = 0$) and the ghost number for
them is $+1$ and $-1$ respectively. This ghost number will also have
relevance with some aspects of differential geometry (see, e.g., Section 5).
Though the conserved and nilpotent charge $Q_{B}$ is the analogue of the
exterior derivative $d$ [7,8], the conserved and nilpotent charge $Q_{AB}$ is
{\it not} the analogue of the co-exterior derivative $\delta$. 
This is due to the fact that the anticommutator between $d$ and $\delta$
is not equal to zero (i.e., $\{ d, \delta \}  \neq 0$) 
whereas $Q_{B}$ and $Q_{AB}$ anticommute
($\{ Q_{B}, Q_{AB} \}  = 0$) with each-other. 
Furhermore, there is no analogue of the Laplacian
operator $\Delta$ in (2.10). This fact can be succinctly expressed as
$$
\begin{array}{lcl}
&& Q_{B}^2 = 0 \qquad d^2 = 0 \qquad
Q_{AB}^2 = 0 \qquad \delta^2 = 0 \nonumber\\
&& \{ Q_{B}, Q_{AB} \} = 0 \qquad  \{ d, \delta \} = \Delta \neq  0.
\end{array} \eqno(2.11)
$$
Recently, it has been pointed out that the cohomologically higher-order
BRST- and anti-BRST operators do not anticommute and their anticommutator
leads to the definition of a higher-order Laplacian operator 
for the compact non-Abelian Lie algebras [36]. This
argument does not apply here in our discussion of 
the Abelian  2-form gauge theory
because here the Lie algebra is a trivial (Abelian) algebra. Furthermore,
we do not consider here the higher-order cohomology discussed in Ref. [36].\\

\noindent
{\bf 3 Dual BRST symmetry}\\

\noindent
In this Section, we  discuss the `dual' BRST symmetry 
which leaves the gauge-fixing term of the Lagrangian density invariant. 
This nilpotent symmetry
should be contrasted with the BRST symmetry (and also anti-BRST symmetry)
where it is the curvature term $ H = d B$, that remains invariant.
Just as one linearizes the gauge fixing term by introducing an 
auxiliary field $B_{\mu}$
and a scalar field $\phi_{1}$ in the case of BRST invaraint
Lagrangian density (2.1), one can linearize the
the kinetic energy term by incorporating another auxiliary field 
${\cal B}_{\mu}$ and 
a different scalar field $\phi_{2}$ 
to obtain the off-shell nilpotent dual BRST 
invariance of the same Lagrangian density
\footnote{ By integrating out the
linearizing field ${\cal B}_{\mu}$ and the scalar field $\phi_{2}$, 
we get back the BRST invariant Lagrangian density (2.1).}. 
Such a BRST- and dual BRST invariant Lagrangian density,
incorporating the above linearizations, is
$$
\begin{array}{lcl}
{\cal L}_{D} &=& \frac{1}{2}\; {\cal B}^\mu {\cal B}_\mu
- \frac{1}{3!} \varepsilon_{\mu\nu\lambda\zeta} {\cal B}^\mu
H^{\nu\lambda\zeta} + {\cal B}^\mu \;\partial_{\mu} \phi_{2} 
- \frac{1}{2} B^\mu B_{\mu} 
+ B_\nu \bigl ( \partial_\mu B^{\mu\nu} - \partial^\nu \phi_{1} \bigr )
\nonumber\\
&-& \partial_\mu \bar \beta \;\partial^\mu  \beta 
+ \bigl ( \partial_\mu \bar C_\nu - \partial_\nu \bar C_\mu \bigr ) \;
\partial^\mu C^\nu + \rho \;\bigl ( \partial_\mu C^\mu + \lambda \bigr )
+ \bigl ( \partial_{\mu} \bar C^{\mu} + \rho \bigr )\; \lambda.
\end{array}\eqno(3.1)
$$
Under the following off-shell nilpotent ($\delta_{D}^2 = 0$) dual BRST 
symmetry transformations:

$$
\begin{array}{lcl}
&&\delta_{D} B_{\mu\nu} = \eta\; \varepsilon_{\mu\nu\lambda\zeta}
\;\partial^\lambda \;\bar C^\zeta \qquad
\delta_{D} \bar C_{\mu} = - \eta \;\partial_{\mu}\; \bar \beta \qquad
\delta_{D} C_{\mu} = \eta \;{\cal B}_{\mu} \nonumber\\
&& \delta_{D} \beta = \eta\;\lambda \;\qquad
\delta_{D} \phi_{2} = - \eta \rho \;\qquad
\delta_{D} \bigl ( \bar \beta, \lambda, \rho, \phi_{1}, B_\mu,
{\cal B}_\mu \bigr ) = 0
\end{array} \eqno(3.2)
$$
the Lagrangian density (3.1) transforms as:
$$
\begin{array}{lcl}
\delta_{D} {\cal L}_{D} = - \eta \;\partial_\mu \;\bigl [\;
\rho {\cal B}^\mu + \lambda \partial^\mu \bar \beta 
+ ( \partial^\mu \bar C^{\nu}  - \partial^\nu \bar C^\mu ) {\cal B}_{\nu} 
\;\bigr ].
\end{array} \eqno(3.3)
$$
Thus, the above Lagrangian density (3.1) remains invariant under the dual BRST
transformations (3.2) and the BRST transformations (2.2) (together with
$\delta_{B} ( {\cal B}_{\mu}, \phi_{2} ) = 0$).
It is appropriate to call
the symmetry transformations (3.2) as the `dual' BRST transformations
because it is the gauge-fixing term (i.e., $ \delta B = 
\partial_{\mu} B^{\mu\nu} dx_{\nu} $: the Hodge dual of the curvature
$ d \;B = H$) of the theory that remains invariant 
and the kinetic energy term (which remains invariant under BRST- 
and anti-BRST symmetries) {\it transforms under it} to compensate
for the transformation of the ghost terms.
The Noether conserved current, derived from the above symmetry
transformations, is:
$$
\begin{array}{lcl}
J^{\alpha}_{D} =  
\varepsilon^{\alpha\beta\rho\sigma} B_{\beta}
\partial_{\rho} \bar C_{\sigma} - {\cal B}^{\alpha} \rho
- \lambda \partial^{\alpha} \bar \beta 
- (\partial^{\alpha} C^{\lambda} - \partial^{\lambda} C^{\alpha}) 
\partial_{\lambda} \bar \beta
- (\partial^{\alpha} \bar C^{\lambda} - \partial^{\lambda} \bar C^{\alpha}) 
{\cal B}_{\lambda}
\end{array} \eqno(3.4)
$$
which ultimately leads to the derivation of a conserved ($\dot Q_{D} = 0$)
and nilpotent ($Q_{D}^2 = 0$) dual BRST charge 
($ Q_{D} = \int d^3 x J^{0}_{D}$) as:
$$
\begin{array}{lcl}
Q_{D} = {\displaystyle \int} d^3 x \; \bigl [\; 
\varepsilon^{0ijk} (B_{i})
\partial_{j} \bar C_{k} - {\cal B}_{0} \rho
- \lambda \dot {\bar \beta} 
- (\partial^{0} C^{i} - \partial^{i} C^{0}) \partial_{i} \bar \beta
- (\partial^{0} \bar C^{i} - \partial^{i} \bar C^{0}) 
{\cal B}_{i}  \;\bigr ] .
\end{array} \eqno(3.5)
$$
To prove the conservation law for the Noether
current in (3.4), one has to use some of the following equations of
motion derived from the Lagrangian density (3.1)
$$
\begin{array}{lcl}
&& \partial \cdot B = 0 \qquad \partial \cdot {\cal B} = 0 \qquad
\Box \phi_{1} = \Box \phi_{2} = 0 \qquad B_{\mu} = \partial^\rho
B_{\rho\mu} - \partial_{\mu} \phi_{1} \nonumber\\
&& {\cal B}_{\mu} = \frac{1}{3!}\; \varepsilon_{\mu\nu\lambda\xi}
H^{\nu\lambda\xi} - \partial_{\mu} \phi_{2} \qquad
\Box \rho = \Box \lambda = \Box \beta = \Box \bar \beta = 0
\nonumber\\
&& \Box C^\mu - \partial^\mu (\partial \cdot C) + \partial^\mu \lambda = 0
\qquad \; \partial_{\mu} C^{\mu} + 2 \lambda = 0, \nonumber\\
&& \Box \bar C^\mu - \partial^\mu (\partial \cdot \bar C)
+ \partial^\mu \rho = 0 \qquad \;
\partial_{\mu} \bar C^\mu + 2 \rho = 0 \nonumber\\
&& \Box B_{\mu} - \partial_{\mu} (\partial \cdot B) = 0 \;\;\rightarrow \;\;
\Box B_{\mu} = 0 \nonumber\\
&& \Box {\cal B}_\mu - \partial_{\mu} (\partial \cdot {\cal B}) = 0
\;\;\rightarrow \;\;\Box {\cal B}_\mu = 0 \nonumber\\
&& \varepsilon_{\mu\nu\lambda\xi} \partial^\lambda {\cal B}^\xi
+ ( \partial_{\mu} B_{\nu} - \partial_{\nu} B_{\mu} ) = 0.
\end{array}\eqno(3.6)
$$

As the ghost part of the Lagrangian density (3.1) remains invariant under 
(2.7), it is very interesting to note that the bosonic
part of this Lagrangian density remains invariant under the following
discrete symmetry transformations
$$
\begin{array}{lcl}
&& {\cal B}_{\mu} \rightarrow \mp i B_{\mu} \qquad
\phi_{2} \rightarrow \mp i \; \phi_{1} \qquad
\phi_{1} \rightarrow \pm i \;\phi_{2} \nonumber\\
&& B_{\mu} \rightarrow \pm i \;{\cal B}_{\mu} \qquad
B_{\mu\nu} \rightarrow \mp \frac{i}{2} \;
\varepsilon_{\mu\nu\lambda\xi}\;B^{\lambda\xi}.
\end{array} \eqno(3.7)
$$
It is straightforward to check that
that the total Lagrangian density (3.1) remains invariant
under the combination of discrete symmetry transformations (2.7) and (3.7).
We note here that the analogue of Hodge
$*$ operation of differential geometry turns out to be the combined symmetries
(2.7) and (3.7). This assertion can be verified by the validity of
the following relation
$$
\begin{array}{lcl}
\delta_{D} \bigl ( \Phi \bigr ) = \pm\; * \; \delta_{B}\; * \; 
\bigl ( \Phi \bigr )
\end{array} \eqno(3.8)
$$
where $(+) -$ stands for the generic field $\Phi$ being (bosonic) fermionic
in nature, $\delta_{D}$ and $\delta_{B}$ are the nilpotent transformations
(2.2) and (3.2) and $*$ operation is the discrete transformations
(2.7) and (3.7). Thus, we note that the dual BRST and BRST variations 
(on a field) are
related to each-other in the same way as the action 
of an exterior derivative $d$ and
co-exterior derivative $\delta = \pm * d * $ on a given differential form.
This symmetry is also reflected in the expressions for BRST- and dual
BRST charges. In fact, it can be readily seen that under the transformations
(2.7) and (3.7), one obtains the following changes for these conserved
and nilpotent charges:
$$
\begin{array}{lcl}
Q_{B} \rightarrow \;\;\; Q_{D} \qquad 
Q_{D} \rightarrow \;\;\; - \;\;Q_{B}.
\end{array} \eqno(3.9)
$$
In the language of symmetry transformations, this fact can be translated 
into: $ \delta_{B} (\Phi) \rightarrow \delta_{D} (\Phi), \quad
\delta_{D} (\Phi) \rightarrow - \delta_{B} (\Phi) $ under (2.7) and (3.7).
Here $\Phi$ is the generic field  representing bosonic as well as fermionic
variables of the theory. It is interesting to note the
similarity between relations (3.9) and the usual electro-magnetic duality
present in the case of Maxwell equations (for $U(1)$ gauge theory) where
$ {\bf E} \rightarrow {\bf B}, \quad {\bf B} \rightarrow - {\bf E}$
under global duality transformations (see, e.g., Ref. [37,38]).

The existence of discrete symmetry for the ghost action, allows one to
define an anti-dual BRST charge $Q_{AD}$ from the expression for $Q_{D}$
in (3.5). The off-shell nilpotent transformations generated by $Q_{AD}$
can be also derived from (3.2) by exploiting (2.7). Now, it is evident 
that the total Lagrangian density (3.1) respects four nilpotent symmetries
which are generated by (anti) BRST- and (anti) dual BRST charges. The 
exact expressions for these charges for the Lagrangian density (3.1) are
$$
\begin{array}{lcl}
Q_{B} = {\displaystyle \int} 
d^3 x \;\bigl [\; \varepsilon^{0ijk} {\cal B}_{i} \partial_{j} C_{k} 
+ B_{0} \lambda
- \rho \dot \beta + (\partial^{0} C^{i} - \partial^{i} C^{0}) B_{i}
- (\partial^{0} \bar C^{i} - \partial^{i} \bar C^{0}) 
\partial_{i} \beta \;\bigr ] 
\end{array}\eqno(3.10) 
$$

$$
\begin{array}{lcl}
Q_{D} = {\displaystyle \int} 
d^3 x \;\bigl [\; 
\varepsilon^{0ijk} B_{i}
\partial_{j} \bar C_{k} - {\cal B}_{0} \rho
- \lambda \dot {\bar \beta} 
- (\partial^{0} C^{i} - \partial^{i} C^{0}) \partial_{i} \bar \beta
- (\partial^{0} \bar C^{i} - \partial^{i} \bar C^{0}) 
{\cal B}_{i}  \;\bigr ] 
\end{array}\eqno(3.11) 
$$

$$
\begin{array}{lcl}
Q_{AB} = i {\displaystyle \int} 
d^3 x \;\bigl [\; \varepsilon^{0ijk} {\cal B}_{i} \partial_{j} \bar C_{k} 
+ B_{0} \rho
+ i \lambda \dot  {\bar \beta} 
+ (\partial^{0} \bar C^{i} - \partial^{i} \bar C^{0}) B_{i}
+ i (\partial^{0}  C^{i} - \partial^{i}  C^{0}) 
\partial_{i} \bar \beta \;\bigr ] 
\end{array}\eqno(3.12) 
$$

$$
\begin{array}{lcl}
Q_{AD} = i {\displaystyle \int} 
d^3 x \;\bigl [\; 
\varepsilon^{0ijk} B_{i} 
\partial_{j}  C_{k} - {\cal B}_{0} \lambda
- i \rho \dot  \beta 
- (\partial^{0} C^{i} - \partial^{i} C^{0}) {\cal B}_{i}
- i (\partial^{0} \bar C^{i} - \partial^{i} \bar C^{0}) 
\partial_{i} \beta  \;\bigr ].
\end{array} \eqno(3.13)
$$

\newpage
\noindent
{\bf 4 Bosonic symmetry}\\

\noindent
It is evident that the total Lagrangian 
density ${\cal L}_{D}$ in (3.1) is endowed
with four nilpotent symmetry transformations that are generated by the
conserved and nilpotent charges (3.10--3.13). It is logical to expect that
the anticommutator of the pair of these symmetries would also be the symmetry
for (3.1). Since four anticommutators ($\{ Q_{B}, Q_{AB} \} = 0,
\{ Q_{D}, Q_{AD} \} = 0, \{Q_B, Q_{AD}\}=0, \{Q_D, Q_{AB}\}=0$) are zero, 
the other two anticommutators 
($\{ Q_{B}, Q_{D} \},
\{ Q_{AB}, Q_{AD} \}$) would lead to the definition of a bosonic operator $W$ 
which will generate a symmetry transformation $\delta_{W}$ for (3.1). The 
following transformations generated by the operator $W$
(with $\kappa = - i \eta \eta^{\prime}$)
$$
\begin{array}{lcl}
&& \delta_{W} B_{\mu\nu} = i \kappa\;
( \partial_{\mu} {\cal B}_{\nu} - \partial_{\nu} {\cal B}_{\mu} 
+ \varepsilon_{\mu\nu\lambda\xi} \partial^\lambda B^\xi )
\quad \delta_{W} \phi_{1} = 0 \quad 
\delta_{W} \phi_{2} = 0, \delta_{W} B_{\mu} = 0 \nonumber\\
&& \delta_{W} C_{\mu} = i \kappa \partial_{\mu} \lambda \qquad
\delta_{W} \bar C_{\mu} = - i \kappa \partial_{\mu} \rho \qquad
\delta_{W} \rho = 0, \quad \delta_{W} \lambda = 0 \qquad
\delta_{W} {\cal B}_{\mu} = 0 \nonumber\\
&& \delta_{W} (\partial \cdot C) = i \kappa \Box \lambda \quad
\delta_{W} (\partial \cdot \bar C) = - i \kappa \Box \rho \quad
\delta_{W} (\partial^\rho B_{\rho\mu}) = i \kappa
(\Box {\cal B}_{\mu} - \partial_{\mu} (\partial \cdot {\cal B})) \nonumber\\
&&\delta_{W} \bigl ( \frac{1}{3!} \varepsilon_{\mu\nu\lambda\xi}
H^{\nu\lambda\xi} \bigr ) \equiv \delta_{W} \bigl ( \frac{1}{2}
\varepsilon_{\mu\nu\lambda\xi} \partial^\nu B^{\lambda\xi} \bigr )
= i \kappa (\Box B_{\mu} - \partial_{\mu} (\partial \cdot B)) \;
\delta_{W} \beta = 0
\; \delta_{W} \bar \beta = 0
\end{array}\eqno(4.1)
$$
turn out to be the symmetry transformations
for ${\cal L}_{D};$ 
$$
\begin{array}{lcl}
\delta_{W} {\cal L}_{D} &=& i \kappa\; \partial_{\alpha} \;[ X^\alpha  ] 
\nonumber\\
 X^\alpha &=& \rho \partial^\alpha \lambda - \partial^\alpha \rho
\lambda + B^\rho \partial^\alpha {\cal B}_\rho \nonumber\\
& - & {\cal B}^\rho
\partial^\alpha B_\rho + {\cal B}^\alpha (\partial \cdot B)
- B^\alpha (\partial \cdot {\cal B}).
\end{array} \eqno(4.2)
$$
Here $\eta$ and $\eta^\prime$ (in the definition of $\kappa$) are the
fermionic spacetime indpendent 
parameters in the transformations corresponding to $\delta_{B}$
and $\delta_{D}$ of eqns. (2.2) and (3.2). The Noether conserved 
current corresponding to the transformations (4.1), is
$$
\begin{array}{lcl}
J^{\alpha}_{W} &=& i  
\varepsilon^{\alpha\beta\rho\sigma} \bigl ({\cal B}_{\beta}
\partial_{\rho} {\cal B}_{\sigma} +  B_\beta \partial_\rho B_\sigma
\bigr ) +  i \partial_{\rho} \bigl ( {\cal B}^\rho B^\alpha
- {\cal B}^{\alpha} B^\rho \bigr )\nonumber\\
 &+& i (\partial^{\alpha} C^{\lambda} - \partial^{\lambda} C^{\alpha}) 
\partial_{\lambda} \rho
+ i (\partial^{\alpha} \bar C^{\mu} - \partial^{\mu} \bar C^{\alpha}) 
\partial_{\mu} \lambda
\end{array} \eqno(4.3)
$$
which finally leads to the derivation of a local conserved charge
($ W = \int d^3x J^{0}_{W}$) as 
$$
\begin{array}{lcl}
W = i {\displaystyle \int} d^3 x \; 
\bigl [\; \varepsilon^{0ijk}\; \bigl (\;
B_{i} \partial_{j}  B_{k} + {\cal B}_{i} \partial_{j} {\cal B}_{k} \bigr )
+ (\partial^{0}  C^{i} - \partial^{i}  C^{0}) 
\partial_{i} \rho 
+ (\partial^{0}  \bar C^{i} - \partial^{i}  \bar C^{0}) 
\partial_{i} \lambda 
\;\bigr ]. 
\end{array} \eqno(4.4)
$$
This charge can be directly computed from the anticommutators of
$\{ Q_{B}, Q_{D} \}$ or $\{Q_{AB}, Q_{AD} \}$ by exploiting the analogue
of canonical (anti)commutators in (2.5) for the Lagrangian density 
(3.1). In fact, all the (anti)commuators of (2.5) remain intact except
the fact that now the canonical momenta w.r.t. $B_{kl}$ becomes
$\varepsilon^{0klm} {\cal B}_{m}$ instead of $H^{0kl}$. Thus, 
one has to replace now $H^{0kl}$ by $\varepsilon^{0klm} {\cal B}_{m}$
in one of the commutators of eqn. (2.5).

There are simpler ways to compute this generator $W$ for the bosonic
symmetry transformations in (4.1). Since the conserved and nilpotent 
charges in (3.10--3.13) are the generators of the nilpotent transformations, 
it can be readily seen that the following equations
$$
\begin{array}{lcl}
\delta_{B} Q_{D} &=& - i \eta \{ Q_{D}, Q_{B} \} = - i \eta W \nonumber\\
\delta_{D} Q_{B} &=& - i \eta \{ Q_{B}, Q_{D} \} = - i \eta W \nonumber\\
\delta_{AB} Q_{AD} &=& - i \eta \{ Q_{AD}, Q_{AB} \} = - i \eta W \nonumber\\
\delta_{AD} Q_{AB} &=& - i \eta \{ Q_{AB}, Q_{AD} \} = - i \eta W 
\end{array}\eqno(4.5)
$$
can be exploited to derive $W$ from the expressions of charges in (3.10--3.13)
and the transformations (2.2) and (3.2). 
It will be noticed that here $\delta_{AB}$
and $\delta_{AD}$ correspond to anti-BRST- and anti-dual BRST transformations
that can be easily derived from eqns. (2.2) and (3.2). 
It is straightforward to check that
$\delta_{D} Q_{B} = - i \eta W$ leads to:
$$
\begin{array}{lcl}
W = i {\displaystyle \int} d^3 x
 \;\bigl [\; \varepsilon^{0ijk} {\cal B}_{i} \partial_{j} {\cal B}_{k} 
+ \rho \dot \lambda
+ (\partial^{0} {\cal B}^{i} - \partial^{i} {\cal B}^{0}) B_{i}
+ (\partial^{0} \bar C^{i} - \partial^{i} \bar C^{0}) 
\partial_{i} \lambda \;\bigr ]. 
\end{array} \eqno(4.6)
$$
We can also obtain an expression for $W$ from the expression for $Q_{D}$
by applying the transformations $\delta_{B}$ 
(i.e., $\delta_{B} Q_{D} = - i \eta W$) as given below
$$
\begin{array}{lcl}
W = i {\displaystyle \int} d^3 x\; \bigl [\; \varepsilon^{0ijk} 
B_{i} \partial_{j}  B_{k} 
+ \lambda \dot \rho
- (\partial^{0}  B^{i} - \partial^{i}  B^{0}) {\cal B}_{i}
+ (\partial^{0}  C^{i} - \partial^{i}  C^{0}) 
\partial_{i} \rho \;\bigr ]. 
\end{array} \eqno(4.7)
$$
It is obvious that the expressions (4.6) and (4.7) bear a different outlook
than the expression derived in (4.4). All these expressions for $W$
are, however, identical if we exploit the appropriate equations of motion.
Similar expressions emerge from the calculations of other expressions
in (4.5).
The most concise form of $W$ that can be derived from (4.5), is
$$
\begin{array}{lcl}
W = i {\displaystyle \int} d^3 x \; 
\bigl [\; \varepsilon^{0ijk}\; \bigl (\;
B_{i} \partial_{j}  B_{k} + {\cal B}_{i} \partial_{j} {\cal B}_{k} \bigr )
+ \lambda\; \dot \rho + \rho \;\dot \lambda  \;\bigr ]. 
\end{array} \eqno(4.8)
$$
It will be noticed that
we have exploited here only the off-shell nilpotent symmetries 
(and conserved charges) for the derivation of $W$.

One important point to be noticed here is the fact
that the operator $W$ 
does not go to zero 
if we exploit the equations of motion. This feature is completely
different from the discussion of the free 2D (non)Abelian gauge theories
in Refs. [18-20] where it has been argued that the topological nature of these
theories is encoded in the vanishing of the operator $W$ when equations
of motion are used and all the fields are assumed to fall off rapidly
at infinity.\\

\noindent
{\bf 5 Extended BRST algebra}\\

\noindent
In this Section, we concentrate on the derivation of an extended
BRST algebra (which is found to be constituted by six conserved charges)
and provide a possible connection of this algebra with the  algebra of the de
Rham cohomology operators of differential geometry. In the normal
BRST algebra, there are three conserved charges (viz., $Q_{g}, Q_{B},
Q_{AB}$) of equations (2.9), (3.10) and (3.12). The existence of
new symmetries, however, provide three more conserved charges
(viz., $Q_{D}, Q_{AD}$ and $W$) which are given by equations (3.11),
(3.13) and (4.8). If one exploits the canonical 
(anti)commutators of equation (2.5) for the Lagrangian density (3.1),
one can show that all the six conserved charges obey the following
extended BRST algebra
$$
\begin{array}{lcl}
&& [ W, Q_{k} ] = 0 \qquad \;\;\;\;k = g, B, AB, D, AD \nonumber\\
&& Q_{B}^2 = 0 \qquad Q_{D}^2 = 0 \qquad Q_{AB}^2 = 0
\qquad Q_{AD}^2 = 0 \nonumber\\
&& i [ Q_{g}, Q_{B} ] = + Q_{B} \qquad i [ Q_{g} , Q_{D} ] = - Q_{D}
\qquad i [ Q_{g}, Q_{AB} ] = - Q_{AB} \nonumber\\
&& i [ Q_{g}, Q_{AD} ] = + Q_{AD} \qquad \{ Q_{B}, Q_{AB} \} = 0 
\qquad \{ Q_{D}, Q_{AD} \} = 0 \nonumber\\
&& \{Q_B, Q_{AD}\}=0 \qquad \{Q_D, Q_{AB}\}=0 \qquad
\{ Q_{B}, Q_{D} \} = \{ Q_{AB}, Q_{AD} \} = W.
\end{array} \eqno(5.1)
$$
A few comments are in order. First of all, it is trivial to see that
the operator $W$ is the Casimir operator for the whole extended BRST
algebra. Secondly, there are four nilpotent (of order two) charges in the
extended BRST algebra. Thirdly, two anticommutators (viz. 
$\{ Q_{B}, Q_{D} \}, \{ Q_{AB}, Q_{AD} \}$) lead to the definition of
of the Casimir operator $W$. And, lastly, the ghost number for
charges $Q_{B}$ and $ Q_{AD} $ is $+1$ and that of $Q_{D}$ and $Q_{AB}$
is $-1$. There are simpler ways to check the validity of the above statements. 
For instance, exploiting the symmetry
transformations of eqns. (2.2), (3.2), (4.1) and an infinitesimal version of
(2.8), it can be easily seen that
$$
\begin{array}{lcl}
&&\delta_{B} W = 0 \;\qquad \delta_{D} W = 0 \;\qquad \delta_{g} W = 0
\nonumber\\
&& \delta_{AB} W = 0 \;\qquad \delta_{AD} W = 0 \;\qquad
\delta_{W} W = 0
\end{array}\eqno(5.2)
$$
where the expression for the $W$ operator can be taken to be its
most concise form of eqn. (4.8).  Similarly other expressions for
the (anti)commutators in (5.1) can be checked by merely using the
symmetry transformation properties and the expressions for the
conserved charges.

Next we present arguments to bring out the analogy between symmetry
generators of this field theoretical model and the de Rham cohomology operators.
It is a well known fact that the de Rham cohomology operators
($ d, \delta, \Delta$) obey the following algebra
$$
\begin{array}{lcl}
&&d^2 = 0 \;\qquad \delta^2 = 0 \;\qquad \Delta = (d + \delta)^2 =
d \delta + \delta d \nonumber\\
&& [ \Delta, d ] = 0 \;\qquad [ \Delta, \delta ] = 0\;
\qquad \;\;\Delta = \{ d, \delta \}.
\end{array}\eqno(5.3)
$$
Furthermore, a differential form of degree $n$ ($ f_{n}$) becomes a
differential form of degree $n + 1$ ($f_{n + 1}$) due to the application 
of operator $d$ (i.e., $ d f_{n} \sim f_{n+1} $). 
In contrast, the operator $\delta$ reduces the degree
of a form by one (i.e., $\delta f_{n} \sim f_{n -1}$) on which it acts
and the Laplacian operator $\Delta$ does not change the degree of
the form (i.e., $ \Delta f_{n} \sim f_{n} $).
Now we observe that the ghost number of the state is parallel to the
degree of the differential form and $Q_B ,Q_D $ and $W$ play respectively 
the role of $d,\delta$ and $\Delta$ in differential geometry.
Exploiting the algebra (5.1), it can be readily
seen that a state $|\psi>_{n}$ with ghost number $n$
(i.e., $ i Q_{g} |\psi>_{n} = n |\psi>_{n}$) in the quantum Hilbert 
space will imply that the ghost number for the states $Q_{B} |\psi>_{n},
Q_{D} |\psi>_{n}, W |\psi>_{n}$ is $ (n+1), (n -1), n $ respectively.
This fact can be succinctly expressed as
$$
\begin{array}{lcl}
i Q_{g} Q_{B} | \psi>_{n} &= & (n + 1)\; Q_{B} | \psi>_{n} \nonumber\\
i Q_{g} Q_{D} |\psi>_{n} &= & (n - 1)\; Q_{D} | \psi>_{n} \nonumber\\
i Q_{g} \;W \;| \psi>_{n} &= & \;n\; W \;| \psi>_{n} \nonumber\\
i Q_{g} Q_{AB} | \psi>_{n} &= & (n - 1)\; Q_{AB} | \psi>_{n} \nonumber\\
i Q_{g} Q_{AD} | \psi>_{n} &= & (n + 1)\; Q_{AD} | \psi>_{n}. 
\end{array}\eqno(5.4)
$$
Thus, now one can draw a parallel between the differential geometry
(and the corresponding de Rham cohomology operators) defined on a compact
manifold and the quantum states, conserved charges, etc., defined in
the quantum Hilbert space of states. For instance, the differential
forms are just like quantum states; a closed form ($d f = 0$) is just like
a BRST closed (physical) state ($Q_{B} |\psi> = 0$); a compact manifold 
is just like the quantum Hilbert space of states; degree of a form is
analogous to the ghost number and the de Rham cohomolgy operators
($d, \delta, \Delta$) have their counterpart as conserved charges
($Q_{B}, Q_{D}, W$) and ($Q_{AD}, Q_{AB}, W$), etc. It is a very
special feature of the BRST formalism that {\it each} of 
the de Rham cohomology operators $d, \delta$ can be identified 
with {\it two} symmetry generators. This, in turn, implies that irrespective
of the nature (i.e., real or complex) of the compact manifold, its
counterpart---- the quantum Hilbert space of states---- is always
complex so that $d$ and $\delta$ have two representations and 
the analogue of the Laplacian operator (i.e., $W$) can also be expressed 
in two different ways (i.e., $ \{ Q_{B}, Q_{D} \} = \{ Q_{AB}, Q_{AD} \} 
= W$). However, if we retrace back, the full strength of the BRST cohomology
and Hodge decomposition theorem implies that the compact manifold has to be 
a complex manifold so that one can achieve a  complete analogy with BRST
formalism. In other words, it should be possible to define ($ d, \bar d$),
($ \delta, \bar \delta$) and ($ \Delta, \bar \Delta$) on the compact manifold
so that cohomology operators $ \Delta = \bar \Delta$ and
$ \Delta = d \delta + \delta d \equiv \bar d \bar \delta + \bar \delta
\bar d$ can be constructed on this manifold.\\

\noindent
{\bf 6 Constraint analysis}\\

\noindent
In this Section, we first discuss the Hodge decomposition theorem for a given
state $|\psi>_{n}$ (with ghost number $n$) in the quantum Hilbert space of
states. This is, then, followed by the discussion of constraints on the
physical (harmonic) states by the imposition of the physicality criteria
with conserved and nilpotent charges (i.e., $ Q_{B} |phys> = 0, 
Q_{D} |phys> = 0$) which define the physical subspace of states in the
total quantum Hilbert space of states. 
It is obvious from the algebra (5.1) and the ghost number analysis
in eqn. (5.4) that, now any arbitrary state $|\psi>_{n}$ in the
quantum Hilbert space of states can be written as
$$
\begin{array}{lcl}
|\psi>_{n} = |\omega>_{n} + Q_{B} |\theta>_{n-1} + Q_{D} |\chi>_{n+1}
\end{array}\eqno(6.1)
$$
where $|\omega>_{n}$ is the harmonic state (i.e., $ W |\omega>_{n} = 0,
Q_{B} |\omega>_{n} = 0, Q_{D} |\omega>_{n} = 0$), $Q_{B} |\theta>_{n-1}$ is
a BRST exact state and $Q_{D} |\chi>_{n+1}$ is a co-BRST exact state. This
equation is just the analogue of the Hodge decomposition theorem (1.1)
written for a differential form in terms of
the de Rham cohomology operators ($d , \delta,  \Delta$)
defined on a compact manifold. It is a special feature of the BRST formalism
(and the corresponding extended BRST algebra (5.1))
that eqn. (1.1) can also be expressed in terms of the conserved and
nilpotent charges $Q_{AB}$ and $Q_{AD}$ as
$$
\begin{array}{lcl}
|\psi>_{n} = |\omega>_{n} + Q_{AD} |\theta>_{n-1} + Q_{AB} |\chi>_{n+1}
\end{array}\eqno(6.2)
$$
where $Q_{AB}$ and $Q_{AD}$ are 
the anti-BRST and anti-dual BRST charges$^{*}$.

\vskip .50cm
\hrule
\vskip.3cm
\noindent {$^{*}$ {\small Unlike the uniqueness of the 
Hodge decomposition in the mathematical
aspects of the de Rham cohomology, the uniqueness of
the corresponding decomposition of the quantum
states (cf. eqns. (6.1, 6.2)) is not obvious in the quantum Hilbert
space of states.}}
\newpage
\noindent
It is a noteworthy point that
the combined discrete transformations (2.7) and (3.7), 
turn out to be  the {\it symmetry} of the Lagrangian density of the theory
under discussion. It is obvious from our earlier arguments that
this symmetry corresponds to the Hodge {\it duality}
operation (i.e., $*$ operation) in differential geometry. 
Thus, we have a theory which is duality invariant due to the
presence of the discrete symmetries (2.7) and (3.7). As
a result, the vacuum- and physical states of the quantum theory should
be also duality (i.e., BRST and dual BRST) invariant in the quantum
Hilbert space of states. This feature, in fact, has been exploited in
Refs. [18-20] to establish the topological nature of the 2D free
(non)Abelian gauge theory. 
In the BRST formalism, physical states are those states that are annihilated 
by $Q_B$ (i.e. $ Q_{B} |phys> = 0$). 
Due to the presence of the discrete symmetry, it
is obvious that $Q_B $ goes to $Q_D$ (cf. (3.9)) and hence the latter also 
annihilate the physical states (i.e., $Q_{B} |phys> = 0 \rightarrow
Q_{D} |phys> = 0$).  These two together 
imply that the Casimir operator $W$ also annhilates 
the physical states. It is, therefore, clear that the physical 
states are the {\it harmonic states}.
Of course, the vacuum state will be annihilated by all 
these charges, as they are the generators of the {\it unbroken} symmetry 
transformations$^{**}$.
Thus, these states satisfy
$$
\begin{array}{lcl}
&& W |vac> = 0 \qquad Q_{B} |vac> = 0 \qquad Q_{D} |vac> = 0\nonumber\\
&& Q_{AB} |vac> = 0 \;\;\;\qquad \;\;\;Q_{AD} |vac> = 0 \nonumber\\
&& W |phys> = 0 \qquad Q_{B} |phys> = 0 \qquad Q_{D} |phys> = 0\nonumber\\
&& Q_{AB} |phys> = 0 \;\;\;\qquad \;\;\;Q_{AD} |phys> = 0. 
\end{array}\eqno(6.3)
$$
The conditions
$i Q_{g} |phys> = 0$ and $ i Q_{g} |vac> = 0$ imply that the ghost number
of the physical- and vacuum states is zero. No other constraint
emerge on physical states due to the existence of ghost charge $Q_{g}$.
It will be noted that the conditions
($Q_{B} |phys> = 0, Q_{AB} |phys> = 0$) lead to the one and the same 
constraints on the physical state. Thus, we can choose one of them for
the constraint analysis. Similar argument holds for the conditions
$Q_{D} |phys> = 0$ and $Q_{AD} |phys> = 0$ and one can choose only
one of these charges for the discussion of constraints$^{\ddagger}$. 
Thus, we see that the vacuum, as well as the physical (harmonic) states,
of the theory respect three {\it basic} symmetries (cf. (6.3)) and the
ghost number for them is zero. It will be noticed that these
conclusions are arrived at by the symmetry considerations {\it alone}.

Before we concentrate on the constraint analysis of the 
Lagrangian density
in (3.1), we shall dwell a bit on the nature of constraints for the original 
Lagrangian density ${\cal L}$ of eqn. (1.2). It is evident that the
canonical momenta w.r.t. the antisymmetric field $B_{\mu\nu}$ is:
\vskip .50cm
\hrule
\vskip .3cm
\noindent {$^{**}$ {\small If the discrete transformations (2.7) and (3.7)
 (which relate $Q_B $and $Q_D$) are not the symmetry of the Lagrangian density,
the physical (harmonic) states can be
assumed to be annihilated independently by the
BRST and the dual BRST charges.}}\\
{$^{\ddagger}$  {\small In what follows, we shall concentrate on the set of
operators $Q_{B}, Q_{D}, W$ for the discussion of 
the Hodge decomposition theorem as well as
the constraint analysis. However, our arguments and analysis will be valid for
the set of operators: $Q_{AB}, Q_{AD}, W$ as well.}}
\newpage
\noindent
$ \Pi^{\mu\nu} = H^{0\mu\nu}$ and the equations of motion are:
$ \partial_{\mu} H^{\mu\nu\lambda} = 0$. Thus, it is clear that
$ \Pi^{0i} \approx 0$ is the primary constraint and the secondary constraint
is nothing but the equation of motion w.r.t. $B_{0i}$ field, i.e.,
$\partial_{j} H^{oij} \equiv \partial_{j} \Pi^{ij} \approx 0$. Both
these constraints are first-class [28,29] in the language of Dirac and they
imply the existence of a gauge symmetry in the theory. 
For the consistent quantization of this theory, it is essential that
$ \Pi^{0i} |phys> = 0,\; \partial_{j} \Pi^{ij} |phys> = 0$ (Dirac's
 presecription). We shall see that exactly
these constraints will appear when we shall demand: $Q_{B} |phys> = 0$ 
(for the Lagrangian density (3.1)). Its dual description will emerge
from the requirement: $Q_{D} |phys> = 0$.

It can be readily seen that the
 requirement $Q_{B} |phys> = 0$, for the Lagrangian
density (3.1), leads to the following constraints on the theory:
$$
\begin{array}{lcl}
\Pi^{0i} (= B^{i}) |phys> = 0 &\;\;\rightarrow\;\;&
(\partial_{\rho} B^{\rho i} - \partial^{i} \phi_{1})
|phys> = 0\nonumber\\
\partial_{j} \Pi^{ij} (= \partial_{0} B^{i}) |phys> = 0 
&\;\;\rightarrow\;\;&
(- \varepsilon^{oijk} \partial_{j} {\cal B}_{k})
|phys> = 0
\end{array}\eqno(6.4)
$$
where the expression for $Q_{B}$ has been taken from eqn. (3.10) and
equations of motion from (3.6) have been used for the above derivation.
Furthermore, it has been assumed here that the ghost fields, present
in the expression for $Q_{B}$, do not lead to any constraints on the
physical states of the theory. It is evident that in the above equation,
we retrieve the constraints of the
original gauge theory described by the Lagrangian density (1.2). Now
the requirement $Q_{D} |phys> = 0$ leads to
$$
\begin{array}{lcl}
({\cal B}^{i}) |phys> = 0 &\;\;\rightarrow\;\;&
(\frac{1}{2} \varepsilon^{i\nu\lambda\xi}
\partial_{\nu} B_{\lambda\xi} - \partial^{i} \phi_{2})
|phys> = 0\nonumber\\
( \partial_{0} {\cal B}^{i}) |phys> = 0 
&\;\;\rightarrow\;\;&
(+ \varepsilon^{oijk} \partial_{j} B_{k})
|phys> = 0.
\end{array}\eqno(6.5)
$$
Exploiting eqn. (3.7) of the duality transformations for the bosonic part
of the Lagrangian density, it can be checked that the above constraints
in (6.5) are just the `dual description'  of the constraints 
obtained in (6.4), though they appear different. 

It will be noticed that even though the auxiliary field $B_{0}$ is 
present in the expression for $Q_{B}$, we have not written 
$Q_{B} |phys> = 0$ implies $B_{0} |phys> = 0$. This is because of the
fact that $B_{0}$ is a conserved quantity and it remains the same
w.r.t. time evolution. In fact, it can be easily seen that the
quantity: $ I_{0} = \int d^3 x B_{0} $ is a time evolution invariant
operator due to
equations of motion in (3.6) (i.e., $\partial_{0} B_{0} = \partial_{i} B_{i}$).
Thus, $B_{0} |phys> = 0$ is a trivial constraint on the theory. Similarly,
we have not concluded from the restriction $Q_{D} |phys> = 0$, the obvious
constraint ${\cal B}_{0} |phys> = 0$ as there is no evolution for the
${\cal B}_{0}$ field due to $ \partial \cdot {\cal B} = 0$ (cf. (3.6)). 
Strictly speaking, however, these constraints should be incorporated
in (6.4) and (6.5) respectively. In fact, these finally imply that
$ \Pi_{\phi_{1}} (= - B_{0}) |phys> = 0 $ and $\Pi_{\phi_{2}} (
= {\cal B}_{0}) |phys> = 0$. More precisely, the constraints
$ B_{0} |phys> = 0, B_{i} |phys> = 0$ and its counterpart
$ {\cal B}_{0} |phys> = 0, {\cal B}_{i} |phys> = 0$ {\it together}
imply that:
$$
\begin{array}{lcl}
({\cal B_{\mu}}) |phys> = 0 &\;\;\rightarrow\;\;&
(\frac{1}{2} \varepsilon^{\mu\nu\lambda\xi}
\partial_{\nu} B_{\lambda\xi} - \partial^{\mu} \phi_{2})
|phys> = 0\nonumber\\
(B_{\mu}) |phys> = 0 &\;\;\rightarrow\;\;&
(\partial_{\rho} B^{\rho \mu} - \partial^{\mu} \phi_{1})
|phys> = 0.
\end{array}\eqno(6.6)
$$
This shows that the total gauge-fixing term $(\partial^\rho B_{\rho\mu}
- \partial_{\mu} \phi_{1}$) and {\it its dual} annihilate the physical states
of the theory. These conditions gauge away some of the degrees of freedom
of the $B_{\mu\nu}$ gauge field. It is straightforward to see that
the constraints $ W |phys> = 0$  does not lead to any new restrictions
on the physical state. In fact, it encompasses both the constraints
given in eqns. (6.4) and (6.5) due to $Q_{B} |phys> = 0$ and
$Q_{D} |phys> = 0$. This is due to the fact that $W= \{ Q_{B}, Q_{D} \}$
and $W|phys> = 0$ implies that 
$Q_{B} |phys> = 0$ and $Q_{D} |phys> = 0$ which are 
in some sense, unique
solutions to the constraint $W |phys> = 0$. It should be recalled that
in the discussion of the de Rham cohomology operators and the Hodge
decomposition theorem, one says that the definition of the harmonic form $h$
($ \Delta h = 0 $) implies that $h$ is closed
($d h = 0$) and co-closed ($\delta h = 0$) together (see, e.g.,
Refs. [9,10]). We note that similar conclusions can be drawn here
from the properties of the 
set of local and conserved charges $W, Q_{B}$ and $Q_{D}$
(or the set $ W, Q_{AD} $ and $Q_{AB}$). \\

\noindent
{\bf 7 Summary and discussion}\\

\noindent
We have shown that the BRST invariant two-form gauge theory in 
four ($3+1$) dimensions
has an additional nilpotent symmetry, called the dual BRST, 
which keeps the gauge 
fixing term invariant. The anti-commutator of both the nilpotent charges
(viz., $Q_B $ and $Q_D$) is the generator ($W$) of a bosonic symmetry
transformation, under which,  the ghost terms 
remain invariant. We can see the parallel between the BRST and the
dual-BRST symmetry: The nilpotent (anti)BRST 
symmetry transformations leave the kinetic 
energy term (more precisely the curvature term) of the free Abelian 
2-form gauge theory invariant. On the other hand, it is the gauge-fixing
term that remains invariant under the (anti)dual BRST symmetry
transformations. Another parallel is: Like the BRST invariant Lagrangian 
density (3.1) can be written 
as the sum of kinetic energy- and the BRST exact terms i.e, ${\cal L}_{KE}
+ \frac{1}{\eta}\;\delta_B (F)$ (where $F$ is a function of the local fields), 
in the same way, we can also express (3.1) as the sum of gauge-fixing 
and the co-BRST exact  parts i.e,
${\cal L}_{GF} + \frac{1}{\eta}\;\delta_D (G)$, namely;
$$
\begin{array}{lcl}
{\cal L}_{D} &=& B_\nu  ( \partial_\mu B^{\mu\nu} -\partial^\nu\phi_{1}) 
+ \frac{1}{ \eta}\;\;\delta_{D} \bigl ( G \bigr ) \nonumber\\
G &=& \frac{1}{2} C_\mu {\cal B}^\mu 
-\frac{1}{6} \varepsilon_{\mu\nu\lambda\sigma}C^\mu H^{\nu\lambda\sigma}
- (\partial^\mu C_\mu + \lambda) \;\phi_{2} 
- (\partial_\mu {\bar C}^\mu + \rho) \; \beta.
\end{array}\eqno{(7.1)}
$$

We have exploited the above symmetries to construct a field theoretical model
 for the Hodge theory
on the four dimensional Minkowskian manifold where all the
de Rham cohomology operators $(d, \delta, \Delta)$ have their 
counterparts as the conserved and nilpotent charges (corresponding to
$d$ and $\delta$) and the bosonic conserved charge $W$ (corresponding
to the Laplacian operator $\Delta$) for the BRST invariant version
of the free 2-form Abelian gauge theory. All these charges are local and
they generate the 
symmetry transformations for the Lagrangian density of this
theory. In the framework of the BRST formalism, it turns out that
the analogue of the Laplacian operator (i.e., $W$) can be represented
in two different ways ($ W = \{ Q_{B}, Q_{D} \} 
= \{ Q_{AB}, Q_{AD} \}$). Thus, $d$ and $\delta$ have two 
representations (i.e., $ d \equiv Q_{B}, Q_{AD},\;
\delta \equiv Q_{D}, Q_{AB}$) in terms of the nilpotent charges.

The  bosonic symmetry generator W  (anticommutator of $Q_B$ and $Q_D$)
turns out to be the
Casimir operator for the extended BRST algebra. Under the transformation,
generated by the Casimir operator, all the fermionic fields either 
do not transform or transform by a vector gauge transformation
(e.g., $\delta_{W} C_{\mu} = i \kappa \partial_{\mu} \lambda, \;
\delta_{W} \bar C_{\mu} = - i \kappa \partial_{\mu} \rho$). It will
also be noticed that all the gauge-fixing terms, for the bosonic as well
as the fermionic fields, transform to the equations of motion under this 
transformations (cf. eqn. (4.1)). 
There exists a discrete symmetry 
transformation in the theory (cf. eqns. (2.7) and (3.7))
which behaves like the analogue of the Hodge $*$ operation. In fact, it relates 
the nilpotent transformations $\delta_D$  and 
$\delta_B$ in a similar fashion as there exists a relationship between
the dual-exterior derivative $\delta$ ($\delta = \pm * d * $) and
the exterior derivative $d$ in differential geometry.

We summarise the main results : (i) 
We have found out a possible mapping between the de Rahm cohomology operators 
of differential geometry and the symmetry 
generators of a $3+1$ dimensional field theoretical model for the Hodge
theory.
(ii) We have shown the existence of a mapping between Hodge $*$ operation
and the discrete transformations on the fields
of the theory. Both these mappings can be concisely expressed as
$$
\begin{array}{lcl}
{\rm Exterior~ derivative}~ d &\;\Leftrightarrow\;\;& Q_{B}, Q_{AD}
     \nonumber\\
{\rm Co-exterior ~derivative}~\delta&\;\Leftrightarrow\;\;& Q_{D}, Q_{AB}
\nonumber\\
{\rm Laplacian} ~\Delta &\;\Leftrightarrow\;\;& W = \{ Q_{B}, Q_{D} \}
= \{ Q_{AB}, Q_{AD} \} \nonumber\\
{\rm Hodge}~ * ~{\rm operation}&\;\Leftrightarrow\;\;& {\rm symmetry~ 
transformations} ~(2.7) ~and ~(3.7).
\end{array}\eqno{(7.2)}
$$
(iii)
The constraints, 
emerging from $Q_{B} |phys> = 0$ and $Q_{D} |phys> = 0$, are related
to each-other due to
the existence of the discrete  duality transformations 
(3.7) for the bosonic
part of the Lagrangian density (3.1).
(iv) We see that the Lapalcian operator $W$
does not go to zero on-shell. This property  was claimed to be
one of the salient features of the topological field theory in
Refs. [18-20] where the topological nature of the free 2D (non)Abelian
1-form gauge theory was established. Furthermore, we are unable
to express the Lagrangian density (3.1) as the sum of BRST- and dual
BRST invariant  parts. This, in turn, implies that the energy-momentum tensor
can also be not expressed as the sum of BRST- and dual BRST 
anticommutators. 
In addition, we are unable to obtain the 
topological invariants of the theory under consideration.
Thus, 2-form free Abelian gauge theory in 4D does not mimic
all the features of the free 2D 1-form gauge theory as a field theoretical
model for the Hodge theory.

It will be interesting to explore the possibility of extending our 
investigations to the case of {\it interacting} (non)Abelian 
two-form gauge theory where matter fields are also present. 
The existence of new symmetries and their generalizations
might turn out to be useful in the context of the proof for the
renormalizability of such theories.
It would be useful if we could discuss the $B \wedge F$ 
(non)Abelian gauge theory, in the
framework of BRST cohomology and Hodge decomposition theorem where
the 1-form gauge fields and 2-form gauge fields are coupled to each-other
in a topologically invariant way. Its further extension to include matter
fields is another workable problem. These are some of the issues which
are under investigations and our results would be reported in our
future publications.
\vskip0.5cm
\newpage
\noindent 
{\bf Acknowledgements}\\

\noindent
One of us (EH) would like to thank the Director, SNBNCBS, Calcutta
for the warm hospitality extended to him during his visit to the Centre
where a part of this work was done.
He also thanks U.G.C., 
Govt. of India, for the financial support through S.R.F. scheme. Fruitful
conversations with A. Lahiri are gratefully acknowledged. EH and MS thank
V. Srinivasan for encouragement.

\end{document}